\documentclass[twocolumn,superscriptaddress,aps,longbibliography,
showpacs,preprintnumbers,amsmath,amssymb,floatfix,nofootinbib]{revtex4-1}
\usepackage{graphicx}
\usepackage{booktabs}
\usepackage{etoolbox}
\AtBeginEnvironment{tabular}{\footnotesize\setlength{\tabcolsep}{3pt}}

\usepackage{newtxtext,newtxmath} 

\usepackage{xcolor}
\usepackage[unicode=true,bookmarks=false,breaklinks=false,
pdfborder={0 0 1},backref=false,
colorlinks=true,citecolor=blue]{hyperref}

\begin{document}
	
\title{Gravitational Wave Detection via Massive-Particle Interferometry}
\thanks{Published in \textit{Scientific Reports} (2026).
	\href{https://doi.org/10.1038/s41598-026-80883-6}
	{doi:10.1038/s41598-026-80883-6}.}

\author{Alireza Maleki}

\affiliation{School of Physics, Institute for Research in Fundamental
	Sciences (IPM), Tehran, Iran}
	
	\begin{abstract}
		Gravitational-wave (GW) observations provide a powerful means of
		testing general relativity and exploring the physics of black holes
		and the early universe, motivating the development of complementary
		detection techniques. Here, we propose and theoretically investigate
		a GW detector based on massive-particle interferometry, in which
		the signal is encoded in the relative phase accumulated by a quantum
		particle traversing a closed-loop trajectory and a reference path.
		We derive the frequency-dependent GW-induced phase response and
		examine the parameters on which it depends. These parameters provide
		a means of tuning the detector response over a broad frequency range,
		motivating the exploration of tabletop detectors for GW signals at frequencies comparable to and below the LIGO band, in the idealized limit. By exploiting the
		trajectory-dependent gravitationally induced phase, the proposed
		scheme offers a distinct and potentially complementary approach to
		GW detection and motivates further investigation of its experimental
		feasibility and practical sensitivity.
	\end{abstract}
	
	\maketitle  

	\section*{Introduction}	
	Gravitational waves (GWs) interact remarkably weakly with matter and have traveled transparently from almost the beginning of the big bang \cite{maggiore2008gravitational1,caprini2020detecting,blanchet2024post}. This peculiar property lays out the bedrock for drawing a complete picture of the universe and opens a new window to the physics of the very early universe \cite{caprini2020detecting,krauss2010primordial}. Being ubiquitous in various fields of study relevant to the test of general relativity, cosmology, astrophysics, and fundamental issues in quantum gravity, the physics of gravitational-wave observations has broad applications \cite{maggiore2008gravitational2,sathyaprakash2009physics}.
	
	As a prominent discovery, the recent detection of gravitational waves in the LIGO interferometer is considered a major milestone in this direction \cite{abbott2016binary}. Analyzing the attained data in the LIGO interferometer has shed new light on some fundamental questions in physics, and it may provide insights into the quantum aspects of black holes \cite{giddings2016gravitational,abedi2017echoes, foit2019testing, maleki2020speed,akhshi2021template,maleki2026area}. However, considering a more general spectrum of GWs, attaining information from the spectrum beyond the scope of the LIGO interferometer is viewed to be extremely important, which has made the demand for new generations of detectors indispensable \cite{maggiore2008gravitational1,sathyaprakash2009physics, maggiore2025comparison,owen2026colloquium, berti2026deci}. Nowadays, ultra-high-precision interferometry, playing a significant role in the quantum framework \cite{giovannetti2011advances,escher2011general,pezze2018quantum,huang2024review}, has highly influenced gravitational-wave detection strategies \cite{schnabel2010quantum}. As a route to ultra-high-precision detection, atomic interferometry has demonstrated remarkable capabilities. Considering the advantage of having very small noise in these detectors, much interest has been ignited in the construction of matter-wave detectors in space \cite{dimopoulos2009gravitational,geiger2017future,abe2021matter,yu202650,junca2026wavefront}. In this regard, techniques developed at the interface of gravity and quantum physics may offer new approaches for precise investigations of the nature of gravity and for understanding broader phenomena in near future \cite{dimopoulos2008general,zhou2015test,maleki2024cosmic}.
	
	In this direction, recent prominent interferometry-assisted proposals demonstrate the capability of gravitationally induced entanglement between two massive particles to serve as a witness of the quantum aspects of gravity \cite{bose2017spin,marletto2017gravitationally,ming2026unveiling,margalit2021realization,maleki2022complementarity,bose2025massive,marletto2025quantum,maleki2026geometric}. To this end, particles with masses of about $10^{-14}$--$10^{-12}\,\mathrm{kg}$ were proposed for the test of quantum gravity \cite{bose2017spin,marletto2017gravitationally,schut2023relaxation,weber2024bose,di2024bose,bose2509spin,maleki2026optimal}. Significant experimental milestones in this direction were recently reported in the realization of interferometry with larger particles, demonstrating spatial superpositions and coherent control necessary for future tests of quantum gravity \cite{margalit2021realization,keil2021stern, pedalino2026probing}. 
	
	Inspired by these studies, we are led to consider the direct interaction of quantum particles and gravitational waves within an interferometric framework. With a suitable choice of mass, this interaction can produce a detectable phase shift, thereby enabling the detection of gravitational waves. The hallmark of this idea is to place quantum particles in a superposition of two distinct paths and exploit the spatially varying tidal potential of the gravitational-wave interaction to accumulate the phase shift. The preparation of a massive particle in such a spatial superposition is essentially the generation of a Schr{\"o}dinger cat state. Given the rapid progress in quantum technologies for preparing large systems in superposition and generating macroscopic Schr{\"o}dinger cat states \cite{arndt2014testing,gerlich2011quantum,bild2023schrodinger,inarrea2025schrodinger}, such a detector may become available in the near future.
	
	Our results suggest that future advances in macroscopic quantum systems could enable the development of tabletop massive-particle detectors. In contrast to kilometer-scale observatories like LIGO, whose design restricts them to a specific frequency band, the characteristic frequency and sensitivity of this interferometric approach are intrinsically determined by tunable experimental parameters. By adjusting the characteristic length scale of the interferometer and the particle speed, the frequency response of the detector can, in principle, be shifted across different regimes. This tunability suggests a potentially complementary pathway to explore frequency bands, such as the low-frequency regime, that are beyond the reach of current gravitational-wave detectors. 
	
	\section*{Detecting gravitational waves using a massive-particle interferometer}
	\label{S2}
	
	Gravitational waves can be described, in the weak-field regime, as small perturbations of the Minkowski spacetime metric,
	\begin{equation}
		g_{\mu\nu}=\eta_{\mu\nu}+h_{\mu\nu},
		\qquad |h_{\mu\nu}|\ll 1,
	\end{equation}
	where $\eta_{\mu\nu}$ is the Minkowski metric and $h_{\mu\nu}$ is the dimensionless metric perturbation. For a plane GW propagating in vacuum, one may adopt the transverse-traceless (TT) gauge, in which $h_{0\mu}^{\rm TT}=0$, $\delta^{ij}h_{ij}^{\rm TT}=0$, and $h_{ij}^{\rm TT}n^j=0$. Here, Latin indices $i,j$ label spatial components, repeated spatial indices are summed over, and $\mathbf{n}$ is the unit vector in the direction of GW propagation. In this gauge, using the metric signature $(-1,+1,+1,+1)$, the line element is
	\begin{equation}
		ds^2=-c^2dt^2+
		\left(\delta_{ij}+h_{ij}^{\rm TT}\right)dx^i dx^j,
		\label{frwGW}
	\end{equation}
	where $c$ is the speed of light and $\delta_{ij}$ is the Kronecker delta. The spatial perturbation can be decomposed into the two independent tensor polarizations permitted by general relativity \cite{maggiore2008gravitational1,sala2026detecting}
	\begin{equation}
		h_{ij}^{\rm TT}
		=h_+e^+_{ij}+h_\times e^\times_{ij},
		\label{hpolar}
	\end{equation}
	where $h_+$ and $h_\times$ are the dimensionless scalar amplitudes of the plus and cross polarizations, respectively, and $e^+_{ij}$ and $e^\times_{ij}$ are the corresponding polarization tensors. They can be constructed from two orthogonal unit vectors $\mathbf p$ and $\mathbf q$, both perpendicular to $\mathbf n$, as $e^+_{ij}=p_i p_j-q_i q_j$ and $e^\times_{ij}=p_i q_j+q_i p_j$ \cite{maggiore2008gravitational1}. For a circular, nonprecessing compact binary, the leading-order quadrupole waveform can be expressed in terms of its two tensor polarizations as \cite{maggiore2008gravitational1,sala2026detecting}
	\begin{align}
		h_+(t)
		&=h(t)\frac{1+\cos^2\iota}{2}
		\cos\Psi(t), \label{eq:hplus}\\
		h_\times(t)
		&=h(t)\cos\iota\,
		\sin\Psi(t), \label{eq:hcross}
	\end{align}
	where \(\iota\) is the inclination angle between the binary's orbital angular-momentum axis and the line of sight, and \(h(t)\) is the characteristic strain-amplitude. The phase \(\Psi(t)\) describes the temporal evolution of the gravitational wave and satisfies
	\begin{equation}
		\Psi(t)=\Psi_0+2\pi\int_{t_0}^{t}f(t')\,\mathrm{d}t',
		\label{eq:gwphase}
	\end{equation}
	where \(f(t)\) is the observed gravitational-wave frequency, \(t_0\) is a reference time, and \(\Psi_0=\Psi(t_0)\) is the initial phase. For a monochromatic signal, whose frequency remains approximately constant over the observation interval, \(f(t)\simeq f\), choosing \(t_0=0\) and \(\Psi_0=0\) gives \(\Psi(t)\simeq \omega t\) where $\omega=2\pi f$ is the GW angular frequency. In this approximation, the amplitude is also treated as constant over the interval considered.
	
	At leading quadrupole order, the characteristic strain amplitude of a cosmological source is approximately \cite{maggiore2008gravitational1}
	\begin{equation}
		h(t)\simeq
		\frac{4\bigl(G\mathcal M_z\bigr)^{5/3}
			\bigl[\pi f(t)\bigr]^{2/3}}
		{c^4d_L},
		\label{eq:gwamplitude}
	\end{equation}
	where \(d_L\) is the luminosity distance, \(G\) is Newton's gravitational constant. The redshifted chirp mass is defined by
	\begin{equation}
		\mathcal M_z=(1+z)\mathcal M,
		\qquad
		\mathcal M=\mu^{3/5}M^{2/5},
		\label{eq:chirpmass}
	\end{equation}
	where \(z\) is the cosmological redshift, \(M=M_a+M_b\) is the source-frame total mass, and	$\mu={M_aM_b}/{M}$ is the source-frame reduced mass. Here, \(M_a\) and \(M_b\) denote the source-frame component masses.
	
	During the inspiral of a quasi-circular binary, gravitational-wave emission carries away orbital energy and angular momentum, causing the orbital semi-major axis (equal to the orbital separation for a circular orbit) to decrease and the orbital and gravitational-wave frequencies to increase, along with the leading-order strain amplitude. As the binary enters the strong-field regime, higher-order corrections become essential, and the inspiral transitions into merger. For a black-hole binary, the remnant subsequently undergoes damped oscillations known as ringdown, which cannot be described by the leading-order quadrupole approximation and require a waveform model covering the merger and ringdown phases \cite{maggiore2008gravitational2,sathyaprakash2009physics}.
	
	For the remainder of this paper, we assume that the gravitational wave propagates along the $z$-axis, perpendicular to the interferometer plane (the $xy$-plane). We work in the long-wavelength regime, in which the GW wavelength is much larger than the characteristic arm length $l$ of the interferometer, $\omega l/c\ll1$. For the tabletop scales considered here ($l\ll1\,\mathrm{km}$), the
	long-wavelength condition $\omega l/c\ll1$ is strongly satisfied
	for known physical gravitational-wave sources. In a local inertial frame, GW effect is described by the relative tidal acceleration of the particles, which can be decomposed into contributions from the two GW polarizations \cite{thorne2000gravitation}
	\begin{equation}
		\begin{aligned}
			\vec{a}_{+}&=\frac{1}{2}\ddot{h}_{+}(x\hat{x}-y\hat{y}),\\ 
			\vec{a}_{\times}&=\frac{1}{2}\ddot{h}_{\times}(y\hat{x}+x\hat{y}).
		\end{aligned}
		\label{accelerationl}
	\end{equation}
	Using these relations, we can define a time-dependent gravitational potential for each polarization as
	\begin{align}
		\Phi_{+}(x,y,t)&=\frac{1}{4}\ddot{h}_{+}({y^2}-{x^2}),
		\label{pchpotentioal1plus} \\
		\Phi_{\times}(x,y,t)&=-\frac{1}{2}\ddot{h}_{\times}(yx).
		\label{pchpotentioal1}
	\end{align}

	For a gravitational wave treated as monochromatic over the
	observation interval, we assume a face-on source ($\iota=0$ in
	Eqs.~\eqref{eq:hplus}--\eqref{eq:hcross}) with initial phase
	$\Psi_0=0$. The plus polarization is then $h_+=h\cos(\omega t)$
	and the cross polarization is $h_\times=h\sin(\omega t)$.
	The scale of the potential is $|\Phi| \sim |h|\omega^2\ell^2/4$, where $\ell$ is the characteristic loop dimension. The weak-field condition $|\Phi|/c^2\ll1$ then requires $\ell \ll c/(\omega\sqrt{|h|})$, which is satisfied by a wide margin given the long-wavelength condition $\omega l/c\ll1$ and $h\ll1$. 
	
\begin{figure}[h]
		\centering
		\includegraphics[scale=0.25]{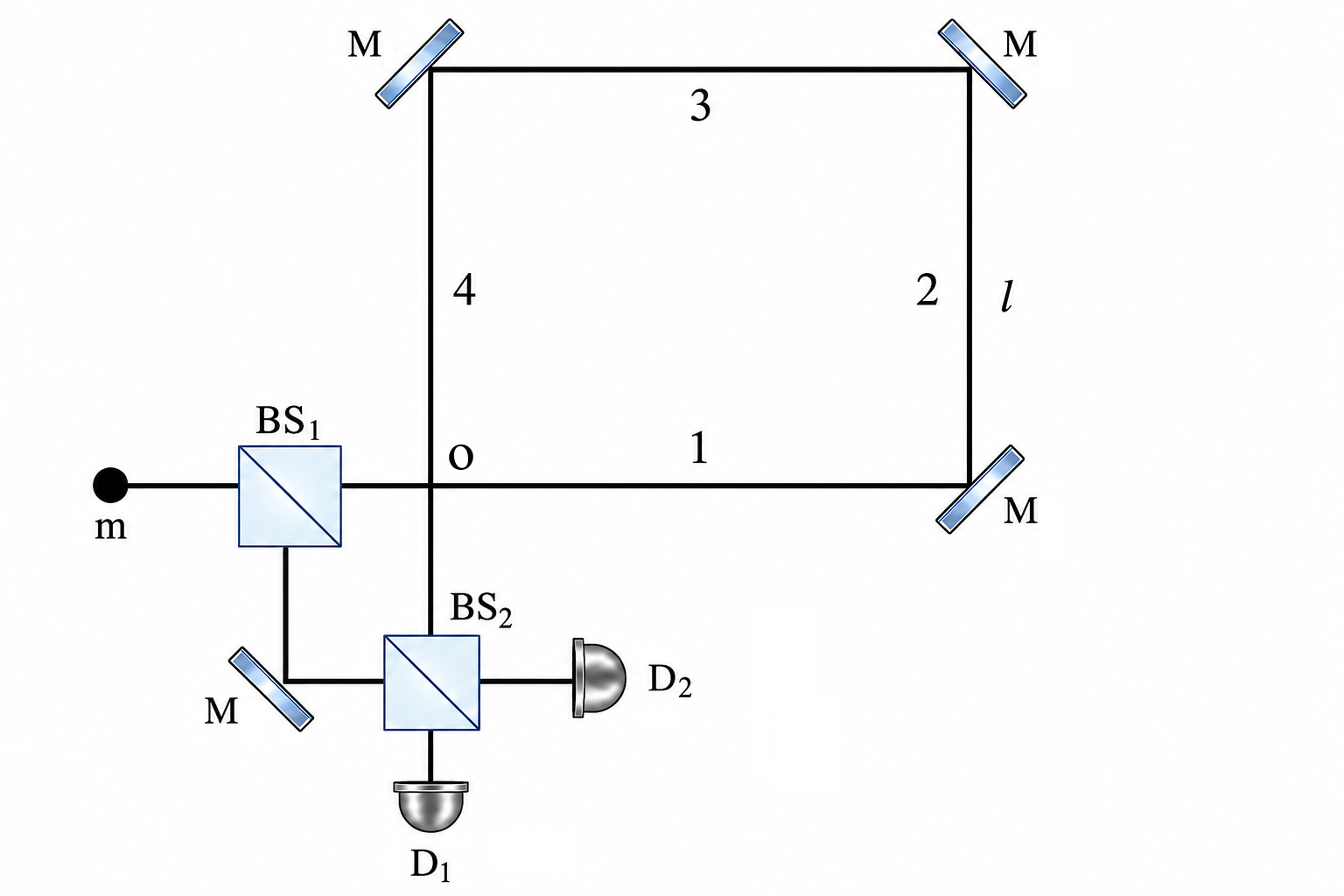}
		\caption{Massive-particle interferometer for GW detection. The particle with mass \(m\) encounters the first beam splitter (BS1) with reflectivity \(r_{1}\) and transmissivity \(t_{1}\). The mirrors are denoted by M, and the length of each arm in the rectangular trajectory is equal to \(l\). After recombining at the second beam splitter (BS2) with characteristics \(r_{2}\) and \(t_{2}\), the interference is detected by the detectors \(D_1\) and \(D_2\). Both beam splitters are assumed to be lossless. The origin of the coordinate system used in the text is denoted by \(O\).}
		\label{MZH}
	\end{figure}

	With these ingredients, we now investigate massive-particle interferometry with a particle of mass $m$, using the setup of Fig.~\ref{MZH}, in which both beam splitters and the detectors are localized near the origin $O$ of the coordinate system. Since the tidal potentials of Eqs.~\eqref{pchpotentioal1plus} and \eqref{pchpotentioal1} vanish at $O$, the reference path connecting the beam splitters accumulates negligible gravitational-wave phase, and the net phase difference arises solely from the loop. The index 1 or 2 in the reflectivity and transmissivity refers to the first and second beam splitters, characterized by $r_i$ and $t_i$ in Fig.~\ref{MZH}, respectively. Both beam splitters are assumed to be lossless, so that \(|r_i|^2 + |t_i|^2 = 1\) holds for \(i=1,2\). 
	
	The particle first encounters the first beam splitter (BS1) with the reflectivity $r_1$ and the transmissivity $t_1$. Therefore, it may be reflected with probability $r_1^2$ toward the detector, taking the lower path right after BS1, or it may be transmitted with probability $t_1^2$ and enter the square loop of the interferometer before being detected. The state immediately after $\mathrm{BS_1}$ is
	\begin{equation}
		\vert \psi_1\rangle
		=r_1\,\vert \psi_{\rm ref}\rangle
		+t_1\,\vert \psi_{\rm loop}\rangle ,
		\label{psiafterBS1}
	\end{equation}
	where $\vert \psi_{\rm ref}\rangle$ and $\vert \psi_{\rm loop}\rangle$ are two orthogonal states associated with the reference arm and the loop, respectively. The loop consists of the four arms labeled $1,2,3,4$ in Fig.~\ref{MZH}. We assume the gravitational waves propagate in the direction of the $z$-axis that is perpendicular to the plane where the interferometer is located. Therefore, Eq.~(\ref{accelerationl}) and its corresponding potential can be applied. The interaction of a particle moving along a path $\mathbf r(t)$ within a gravitational field $\Phi[\mathbf r(t),t]$ results in an induced phase given by \cite{colella1975observation}
	\begin{equation}
		\phi[\mathbf r]=-\frac{m}{\hbar}\int\Phi[\mathbf r(t),t]\mathrm{d}t,
		\label{phasepotential}
	\end{equation}
	where \(m\) denotes the mass of the particle. As the particle traverses the loop, its wave function accumulates a gravitational-wave phase
	\begin{equation}
		\delta\phi_{\rm GW}=\sum_{j=1}^{4}\delta\phi_j ,
		\label{deltaphiGW}
	\end{equation}
	where $\delta\phi_j$ is the phase induced by the gravitational wave in arm $j$. Hence, right before reaching $\mathrm{BS_2}$, the loop component of the state carries the additional phase factor $e^{-i\delta\phi_{\rm GW}}$,
	\begin{equation}
		\vert \psi_2\rangle
		=r_1\,\vert \psi_{\rm ref}\rangle
		+t_1\,e^{-i\delta\phi_{\rm GW}}\,\vert \psi_{\rm loop}\rangle .
		\label{psibeforeBS2}
	\end{equation}
	
	The two components now recombine at $\mathrm{BS_2}$, whose amplitude reflectivity and transmissivity are $r_2$ and $t_2$. We adopt the standard convention in which the beam splitter acts on the incoming states as
	\begin{align}
		\vert \psi_{\rm ref}\rangle
		&\;\longrightarrow\;
		r_2\,\vert \psi \rangle_{D_2}
		+t_2\,\vert \psi \rangle_{D_1} ,
		\label{BSref} \\
		\vert \psi_{\rm loop}\rangle
		&\;\longrightarrow\;
		t_2\,\vert \psi \rangle_{D_2}
		-r_2\,\vert \psi \rangle_{D_1} ,
		\label{BSloop}
	\end{align}
	where the minus sign in Eq.~\eqref{BSloop} guarantees the orthogonality of the basis. Substituting Eqs.~\eqref{BSref} and~\eqref{BSloop} into Eq.~\eqref{psibeforeBS2} and collecting the coefficients of $\vert \psi \rangle_{D_1}$ and $\vert \psi \rangle_{D_2}$ separately, we obtain the outgoing state
	\begin{align}
		\vert \psi \rangle
		&=\bigl(r_1 t_2-r_2 t_1\,e^{-i\delta\phi_{\rm GW}}\bigr)
		\vert \psi \rangle_{D_1} \\
		&+\bigl(r_1 r_2+t_1 t_2\,e^{-i\delta\phi_{\rm GW}}\bigr)
		\vert \psi \rangle_{D_2} \nonumber,
		\label{psifinal}
	\end{align}
	where the states $\vert \psi \rangle_{D_1}$ and $\vert \psi \rangle_{D_2}$ form an orthogonal basis for the output ports of $\mathrm{BS_2}$.
	
	The detection probabilities are obtained by projecting Eq.~\eqref{psifinal} onto each output state. For balanced $50{:}50$ beam splitters, $r_i=t_i=1/\sqrt{2}$, one finds
	\begin{align}
		P_{D_1}
		&=\bigl|\langle \psi \vert \psi \rangle_{D_1}\bigr|^2
		=\frac{1}{2}\bigl(1-\cos\delta\phi_{\rm GW}\bigr) \\
		P_{D_2}
		&=\bigl|\langle \psi \vert \psi \rangle_{D_2}\bigr|^2
		=\frac{1}{2}\bigl(1+\cos\delta\phi_{\rm GW}\bigr) ,
		\label{PD2}
	\end{align}
	We now compute the phase induced by the gravitational wave on the particle. We assume that the particle traverses the four straight arms of the square loop with a constant speed \(v\). For the plus polarization, assuming the strain oscillates harmonically in time at angular frequency \(\omega\), its second time derivative is simply \(\ddot{h}_{+}(t)=-\omega^{2}h\cos(\omega t)\). Inserting this into the tidal potential of Eq.~\eqref{pchpotentioal1plus} and evaluating the phase integral of Eq.~\eqref{phasepotential} along each of the four straight segments of the loop, we find the induced phase of the plus-polarization wave in each arm to be
	\begin{equation}
		\begin{aligned}
			\delta\phi_{1}&=-\frac{mh\omega^2}{4\hbar} \int_{0}^{l/v} \cos(\omega t)\, v^2t^2 \,\mathrm{d}t,\\
			\delta\phi_{2}&=\frac{mh\omega^2}{4\hbar} \int_{l/v}^{2l/v} \cos(\omega t)\left[v^2(t-l/v)^2-l^2\right]\,\mathrm{d}t,\\
			\delta\phi_{3}&=\frac{mh\omega^2}{4\hbar} \int_{2l/v}^{3l/v} \cos(\omega t)\left[l^2-v^2(3l/v-t)^2\right]\,\mathrm{d}t,\\
			\delta\phi_{4}&=\frac{mh\omega^2}{4\hbar} \int_{3l/v}^{4l/v}\cos(\omega t)\, v^2(4l/v-t)^2 \,\mathrm{d}t.
		\end{aligned}
		\label{deltaph1i}
	\end{equation}
	Evaluating these integrals and summing the contributions from all four arms yields the phase induced by the plus polarization as
	\begin{equation}
		\delta \phi_{GW}^{+}
		=\frac{m h v^{2}}{\hbar \omega}\,\mathcal{F}^{+}(\xi) ,
		\label{deltaphi}
	\end{equation}
	where the dimensionless response function is given by
	\begin{equation}
		\mathcal{F}^{+}(\xi)
		=2
		\cos\xi\,\bigl(1-\cos\xi\bigr)
		\bigl(\xi-\sin 2\xi+\xi\cos\xi\bigr)
		,
		\label{Fplus}
	\end{equation}
	and $\xi=\omega l/v$. This result shows that the controllable parameters in massive-particle interferometry are the arm lengths, the particle speed, and the mass of the quantum system. 
	
	Applying the same procedure for the cross polarization, we find
	\begin{equation}
		\delta \phi_{GW}^{\times}
		=\frac{m h v^{2}}{\hbar \omega}\,\mathcal{F}^{\times}(\xi) ,
		\label{deltaphicross}
	\end{equation}
	with the corresponding response function
	\begin{equation}
		\mathcal{F}^{\times}(\xi)
		=\xi\,\sin 2\xi\ \bigl(1-\cos\xi\bigr) .
		\label{Fcross}
	\end{equation}
	Since the total accumulated phase is the sum of these two contributions, we can write
	\begin{equation}
		\delta \phi_{GW}
		=\delta \phi_{GW}^{+}+\delta \phi_{GW}^{\times}=\frac{m h v^{2}}{\hbar \omega}\mathcal{F}(\xi),
		\label{phasetotal}
	\end{equation}
	where we define $\mathcal{F}(\xi)=\mathcal{F}^{+}(\xi)+\mathcal{F}^{\times}(\xi)$. 
	In a single-particle interferometer, a phase shift of \(\pi\) corresponds to a complete inversion of the interference fringe (from constructive to destructive). Setting this phase shift (\(\delta\phi_{\rm GW}=\pi\)) means \(h_{\min}\) denotes the strain amplitude required to produce this complete fringe inversion. The minimum detectable strain is therefore defined for each polarization individually as
	\begin{equation}
		h_{\min}^{A}(\omega)
		=\frac{\pi\,\hbar\,\omega}
		{m\,v^{2}\,\bigl|\mathcal{F}^{A}(\xi)\bigr|} ,
		\qquad A\in\{+,\times\} ,
		\label{hminA}
	\end{equation}
	and for the combined response as
	\begin{equation}
		h_{\min}^{\rm tot}(\omega)
		=\frac{\pi\,\hbar\,\omega}
		{m\,v^{2}\,
			\bigl|\mathcal{F}(\xi)\bigr|} .
		\label{hmintot}
	\end{equation}
	
	\begin{figure}[!ht]
		\centering
		\includegraphics[width=01\linewidth]{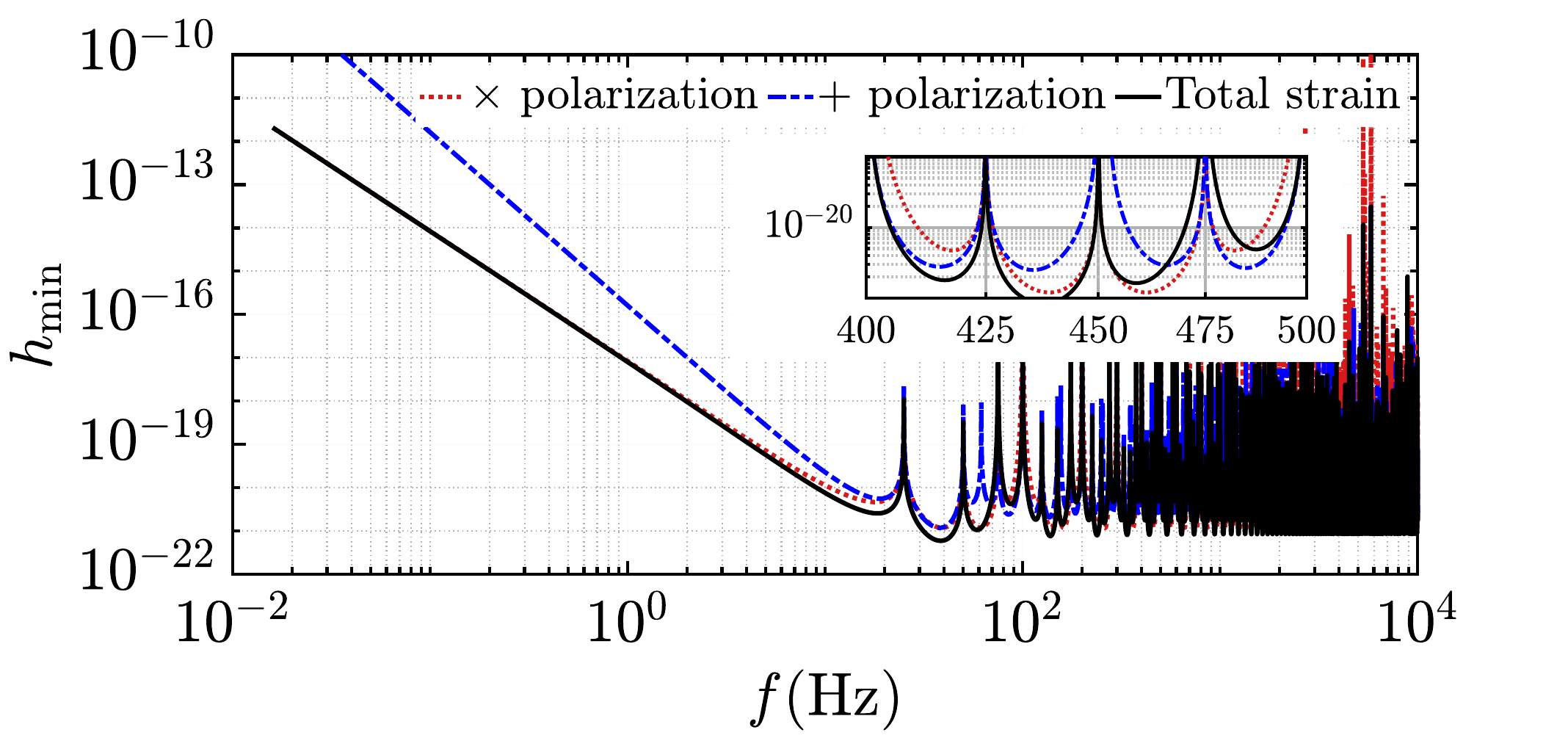}
		\caption{Minimum detectable strain $h_{\min}$ as a function of frequency for the massive-particle interferometer with $m=10^{-14}\,\mathrm{kg}$, $v=100\,\mathrm{m\,s^{-1}}$, and $l=1\,\mathrm{m}$. The dash-dotted curve is the plus polarization $h_{\min}^{+}$, and the dotted curve is the cross polarization $h_{\min}^{\times}$. The solid curve is the total response \(h_{\min}^{\rm tot}\). The inset shows a magnified view of the region $\xi>1$, where the	individual response functions vanish at isolated frequencies.}
		\label{SAGW}
	\end{figure}
	
	\begin{figure}[!ht]
		\centering
		\includegraphics[width=01\linewidth]{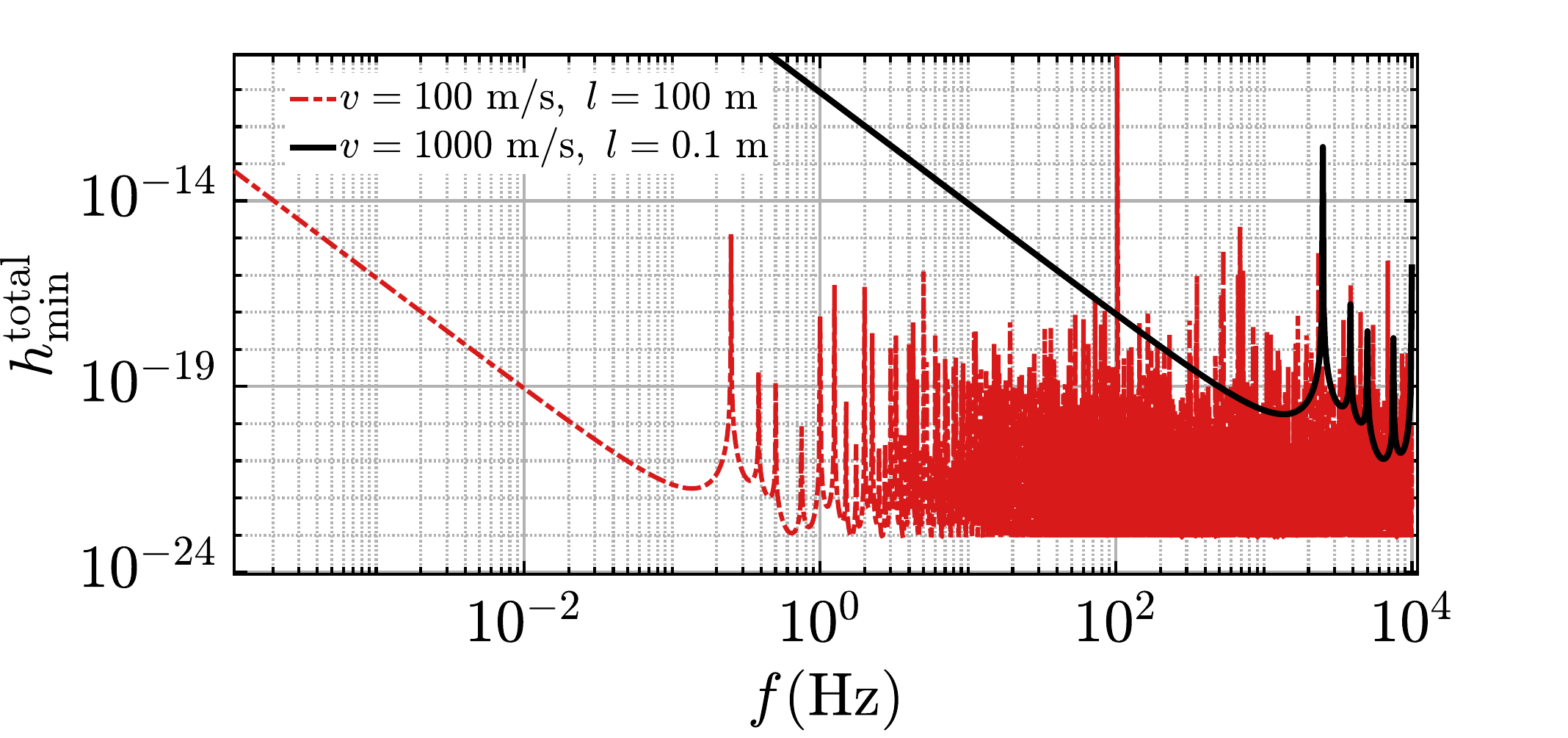}
		\caption{Minimum detectable strain $h_{\min}^{\rm total}$ as a function of frequency for the massive-particle interferometer with $m=10^{-14}\,\mathrm{kg}$, for two parameter sets. The solid  curve corresponds to $v=1000\,\mathrm{m\,s^{-1}}$ and $l=0.1\,\mathrm{m}$, and the dash-dotted  curve corresponds to $v=100\,\mathrm{m\,s^{-1}}$ and $l=100\,\mathrm{m}$.}
		\label{diffvSAGW}
	\end{figure}
	
	The resulting curves from Eqs.~\eqref{hminA} and~\eqref{hmintot} are plotted in Figs.~\ref{SAGW} and~\ref{diffvSAGW}. \\
	Fig.~\ref{SAGW} shows $h_{\min}$ as a function of frequency for $m=10^{-14}\,\mathrm{kg}$, $v=100\,\mathrm{m\,s^{-1}}$, and $l=1\,\mathrm{m}$. Three curves are plotted: the dash-dotted curve is the plus polarization $h_{\min}^{+}$, the dotted curve is the cross polarization $h_{\min}^{\times}$, and black curve is the total strain $h_{\min}^{\rm tot}$.  As can be seen, in our assumed setup the cross polarization can accumulate more induced phase than the plus polarization in the low-frequency regime; therefore, a smaller strain results in a detectable signal. For large frequencies, $\xi>1$, there are isolated frequencies where the response functions vanish; at those frequencies the individual curves diverge. As shown in the inset of Fig.~\ref{SAGW}, these divergences occupy a small fraction of the frequency band, so the detector remains useful in the surrounding range.\\	
	We can adjust the parameters to enhance the response and to extend the detection band. In Fig.~\ref{diffvSAGW} we show $h_{\min}^{\rm total}$ for two parameter choices that differ both in the arm length and in the particle speed. The solid  curve corresponds to a short loop, $l=0.1\,\mathrm{m}$, with fast particles, $v=1000\,\mathrm{m\,s^{-1}}$. The dash-dotted curve corresponds to a long loop, $l=100\,\mathrm{m}$, with slower particles, $v=100\,\mathrm{m\,s^{-1}}$. The two curves illustrate the two distinct regimes of the detector. The long-loop, slow-particle configuration reaches its best response at low frequencies, near $f\sim10^{-1}\,\mathrm{Hz}$, and is therefore suited to probing the infrasound band. The short-loop, fast-particle configuration maintains a smooth response over a broad band and extends the reach toward higher frequencies. Combining the two regimes would in principle allow coverage of a wide frequency window, from below $0.01\,\mathrm{Hz}$ up to the kilohertz range. Since the LIGO detector operates between $10\,\mathrm{Hz}$ and $10\,\mathrm{kHz}$, this  setup can extend the accessible band well below the LIGO range in a tabletop instrument. \\
	To isolate the gravitational-wave signal, a practical strategy is to calibrate 
	and subtract the static phase offset. This offset typically arises from the 
	Earth's gravitational field, static local mass distributions, and instrumental 
	imperfections, and can be calibrated in the absence of a signal (or averaged 
	over a period much longer than the expected signal duration). The remaining 
	time-dependent phase, $\delta\phi_{\rm GW}(t)$, can then be extracted as the 
	gravitational-wave signal.
	To enhance the measurement precision, one can send a burst of \(N\) particles, rather than a single particle, through the interferometer per trigger, with the triggers repeated at a rate \(R\). Since each particle yields an independent measurement of the interference phase, the phase uncertainty per trigger scales as \(\sigma_\phi = 1/\sqrt{N}\). Consequently, the minimum detectable strain is reduced as \(h_{\min} \propto 1/\sqrt{N}\) \cite{giovannetti2011advances,maleki2023speed}. Furthermore, the trigger rate \(R\) defines the sampling rate of the reconstructed signal. Therefore, increasing \(N\) improves the strain sensitivity, while increasing \(R\) enhances the temporal resolution of the detector.
	
	\begin{figure}[!htbp]
		\centering
		\includegraphics[width=1\linewidth]{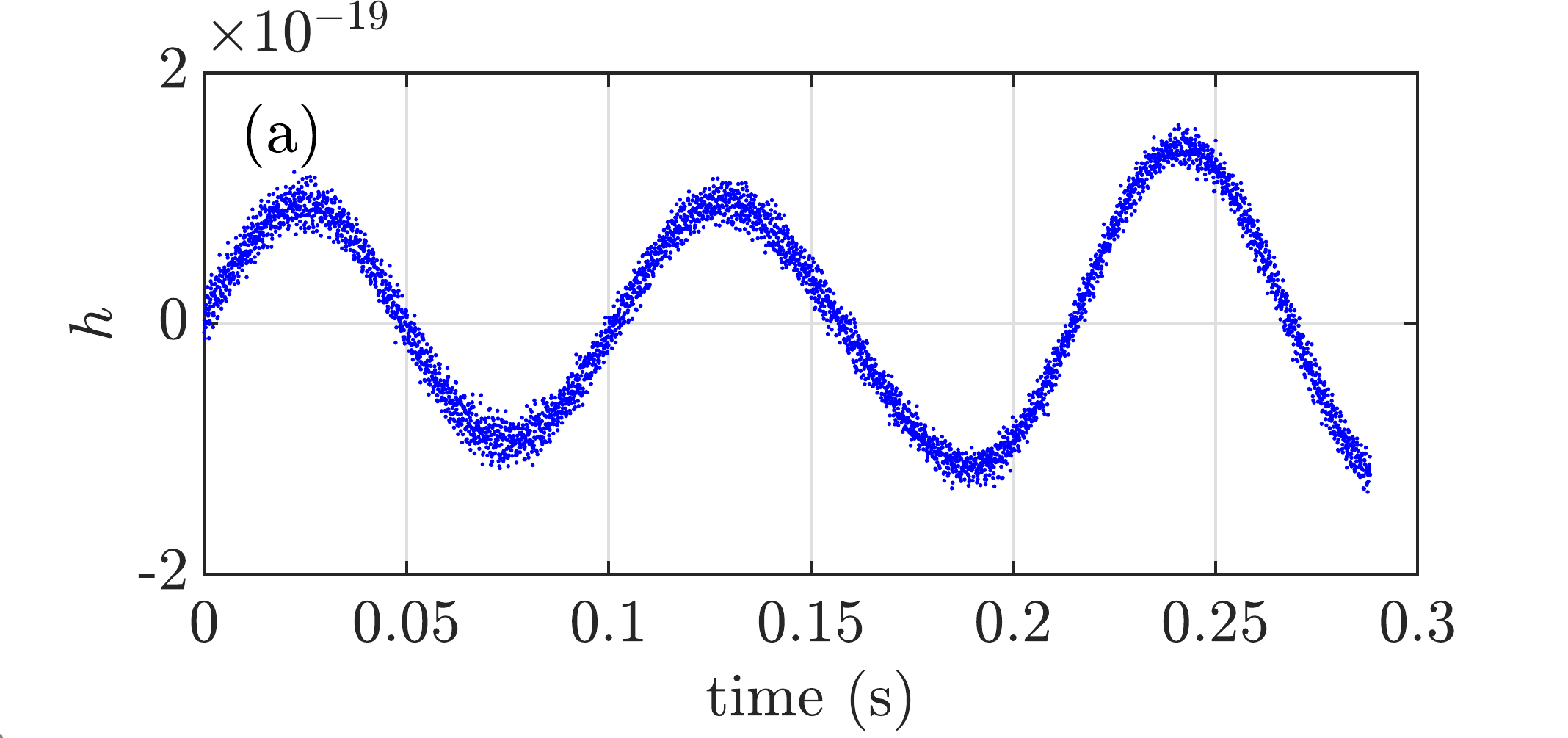}
		\includegraphics[width=1\linewidth]{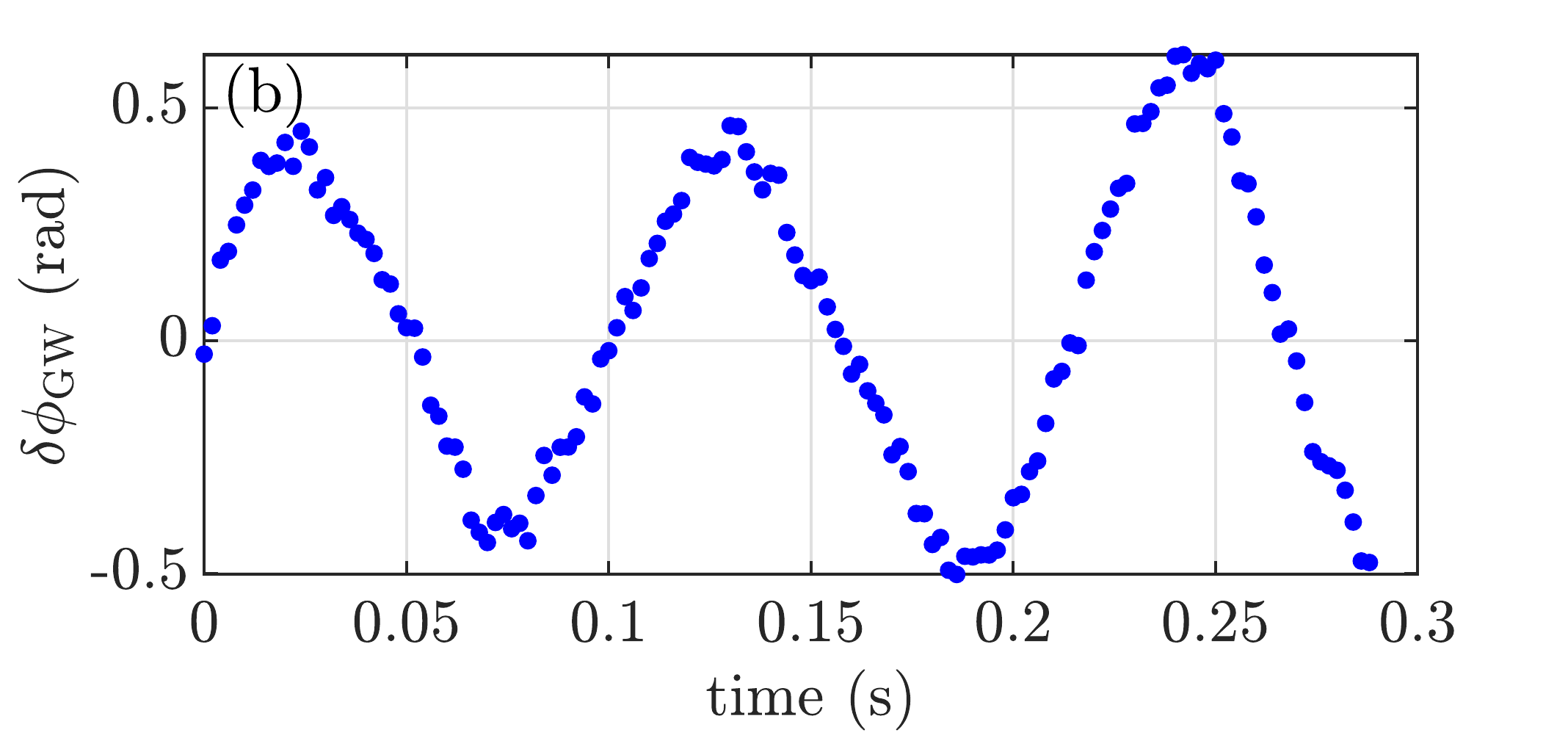}
		\caption{(a) The input gravitational-wave strain time series for the event GW190521 as observed by the LIGO Hanford detector \cite{abbott2020gw190521}. The source is a binary black hole system with component masses of about $85M_{\odot}$ and $66M_{\odot}$ at a luminosity distance of $d_L\sim 5.3\,\mathrm{Gpc}$, rotating with a frequency of about $10\,\mathrm{Hz}$. (b) The simulated output response of massive-particle interferometer driven by this same strain. The accumulated phase $\delta\phi_{\rm GW}(t)$ is plotted for a particle of mass $m=10^{-16}\,\mathrm{kg}$, speed $v=100\,\mathrm{m/s}$, and  length $l=1\,\mathrm{m}$. Each point represents a simulated measurement sampled at $500$ triggers per second. The data in panel (a) is obtained from Ref.~\cite{LIGO_P2000020}.}
		\label{hsamplePhase}
	\end{figure}	
	We now simulate the realistic detection scenario by performing a forward 
	simulation. We take $v = 100\,\mathrm{m/s}$ and $l = 1\,\mathrm{m}$. 
	Figure~\ref{hsamplePhase}(a) shows the input strain time series of GW190521 
	as observed by the LIGO Hanford detector \cite{abbott2020gw190521}. The source 
	is a binary black hole with component masses of about $85M_{\odot}$ and 
	$66M_{\odot}$ at a luminosity distance $d_L \sim 5.3\,\mathrm{Gpc}$, and a 
	gravitational-wave frequency of about $10\,\mathrm{Hz}$. Using this strain as 
	the input signal, Figure~\ref{hsamplePhase}(b) shows the simulated phase 
	accumulated by massive-particle interferometer, sampled at $500$ triggers per second. Each 
	point represents a simulated measurement from one trigger. The particle mass 
	is taken to be $m = 10^{-16}\,\mathrm{kg}$, indicating that even lighter 
	particles can produce a measurable phase accumulation for this source.
	
	\section*{Experimental challenges of gravitational wave detection with massive-particle interferometer}
	\label{S3}
	In this section, we evaluate the practical feasibility of the proposed detector by analyzing potential sources of error and noise. To establish conservative bounds, we first examine the low-frequency regime, where the system is most sensitive to parametric uncertainties. We then estimate the phase shifts induced by environmental factors, such as air molecules and dust, followed by an analysis of the precision required for the experimental parameters (mass, loop size, and particle speed). 
	
	We begin by expanding the response function of Eq.~(\ref{phasetotal}) in the low-frequency limit ($\xi \ll 1$). The total accumulated phase at leading order is given by
	\begin{align}
		\bigl|\delta \phi_{GW}\bigr|
		\approx \frac{m h v^2}{\hbar \omega}\,\xi^4
		= \frac{m h \omega^3 l^4}{\hbar v^2}.
		\label{approxDeltaphi2}
	\end{align}
	Because the induced phase grows with frequency before reaching a flat envelope, the conservative bounds derived in this section for the low-frequency regime remain valid and become even stronger at higher frequencies.
	
	As discussed earlier, the promising advances in quantum interferometry with macroscopic particles are central to our proposal. Present experiments already control quantum systems consisting of $10^{12}$ atoms \cite{leggett1980macroscopic,julsgaard2001experimental, lee2011entangling,klimov2015quantum,ockeloen2018stabilized}, indicating that particles with masses of order $10^{-14}\,\mathrm{kg}$ appear attainable in the near future. While the preparation and coherent control of such massive systems currently remain the most challenging experimental hurdle, progress in this field is continuous. Promising candidate systems include massive molecules \cite{arndt1999wave,eibenberger2013matter,fein2019quantum}, Bose-Einstein condensates \cite{muntinga2013interferometry,proukakis2025century}, and microcrystals \cite{bose2017spin,marshman2022constructing,bose2025massive}.
	
	Beyond the challenge of preparing the quantum state, the detector must also operate in an environment where the GW-induced phase shift is not masked by background noise. We first estimate the phase that may be induced by surrounding particles through their gravitational interaction with the system. Although this background can, in principle, be averaged out, it is still important to evaluate the magnitude of this effect. Let $m_a$ denote the average mass of a surrounding molecule. The atmosphere consists of roughly $78\%$ nitrogen with molecular mass $4\times10^{-26}\,\mathrm{kg}$ and $21\%$ oxygen with molecular mass $2.5\times10^{-26}\,\mathrm{kg}$ \cite{lutgens1995atmosphere}. Taking $m_a \sim 10^{-25}\,\mathrm{kg}$ is therefore a conservative estimate for the most effective molecules in our proposal. For a gas with molecular number density $n$, the characteristic intermolecular separation is $d\sim n^{-1/3}$. Approximating a molecule at this distance as a point mass, the magnitude of its Newtonian interaction energy with the interferometer particle is $|U_a|\sim Gmm_a/d$. The accumulated phase is obtained by integrating the interaction energy over the transit time, $\tau\simeq4l/v$. To estimate the characteristic scale of the interferometric phase perturbation, we assume that the interaction energy along the loop is of the same order as \(\lvert U_a\rvert\). This order-of-magnitude estimate yields
	\begin{equation}
		\begin{aligned}
			|\delta\phi_a|\sim \frac{|U_a|\tau}{\hbar}
			\sim G\frac{mm_a}{\hbar d}\frac{4l}{v}.
		\end{aligned}
		\label{AIRapprox}
	\end{equation}	
	The ratio of this environmental phase to the accumulated low-frequency GW phase from Eq.~(\ref{approxDeltaphi2}) is therefore
	\begin{equation}
		\bigl|\frac{\delta \phi_{a}}
		{\delta \phi_{GW}}\bigr|
		\approx
		\frac{4\,G\,m_av}
		{\omega^{3}\,l^{3}\,d\,h}.
		\label{fracAIRapprox}
	\end{equation}
	This ratio decreases with increasing detector size $l$ and with increasing frequency $\omega$. For the GW190521 event, taking a representative strain $h_0 \simeq 10^{-19}$, together with $m_a \sim 10^{-25}\,\mathrm{kg}$, $l = 1\,\mathrm{m}$, and $v = 100\,\mathrm{m/s}$, Eq.~(\ref{fracAIRapprox}) gives a ratio of order $10^{-11}$ at the atmospheric density $n \simeq 2.7 \times 10^{19}\,\mathrm{cm^{-3}}$. The condition $|\delta\phi_a / \delta\phi_{\rm GW}| \ll 1$ is therefore overwhelmingly satisfied, and the effect of surrounding molecules is negligible. The margin improves further for larger detectors, where the ratio falls as $l^{-3}$, and at higher frequencies, where it falls as $\omega^{-3}$. This setup therefore does not require a high-vacuum environment for the ranges considered here.
	
	We also consider the dust contribution. In a typical urban environment, the air contains about $3.5 \times 10^{7}$ particles per cubic meter with sizes of $0.5\,\mu$m or larger, corresponding to an ISO~9 cleanroom. In an ISO~1 cleanroom, the count drops to about $12$ particles per cubic meter \cite{bhatia2012hvac}. Taking the conservative ISO~9 scenario, the dust mass bound for which the dust phase remains below the GW phase is of order $3 \times 10^{-9}\,\mathrm{kg}$ for $l = 1\,\mathrm{m}$. A typical \(0.5\,\mu\mathrm{m}\) dust particle has a mass of about \(6 \times 10^{-17}\,\mathrm{kg}\), which is roughly seven to eight orders of magnitude smaller than the obtained bound. Dust is therefore not a limiting concern in this configuration. For detectors with arm lengths on the order of \(l = 100\,\mathrm{m}\) at fixed speed, the bound relaxes by a further factor of \(10^{6}\), so the dust mass is correspondingly even farther below the limit. \\
	In addition to environmental noise, the practical feasibility of the detector depends on the operational time scales and the preservation of quantum coherence. The time scale $4l/v$, which sets the experimental cycle of the proposal, is of order a few seconds. This is within reach of current interferometric experiments \cite{margalit2021realization,rosi2014precision}. Finally, large quantum systems are fragile against decoherence, so the isolated path traversed by the particle must be maintained at low temperature and low pressure. For a particle of mass $10^{-14}\,\mathrm{kg}$, preserving the quantum superposition requires pressures on the order of $10^{-15}\,\mathrm{Pa}$ and temperatures below about $0.1\,\mathrm{K}$ \cite{bose2017spin}. Such conditions are attainable with current cryogenic and vacuum technologies \cite{ishimaru1989ultimate}.
	
	The precision of the detector is also constrained by how accurately we can adjust its defining parameters. Specifically, the particle mass \(m\), loop dimension \(l\), and speed \(v\) are each subject to finite measurement precision. Differentiating Eq.~(\ref{phasetotal}) gives the conservative bound
	\begin{equation}
		\left|\frac{\Delta(\delta\phi_{\rm GW})}{\delta\phi_{\rm GW}}\right|
		\;\le\;
		\frac{\Delta m}{m}
		+ \left|\frac{\xi\,\mathcal F'}{\mathcal F}\right|\frac{\Delta l}{l}
		+ \left|2 - \frac{\xi\,\mathcal F'}{\mathcal F}\right|\frac{\Delta v}{v},
		\label{deltaparam}
	\end{equation}
	The three coefficients follow from the structure of Eq.~(\ref{phasetotal}). The mass appears only in the prefactor and is therefore linear, with \(c_m=1\). The length enters only through \(\xi\), so its coefficient is the logarithmic derivative of \(\mathcal F\) and approaches \(c_l\simeq4\) at low frequency, consistent with the \(l^4\) scaling of Eq.~(\ref{approxDeltaphi2}). Conversely, in the high-frequency region where the minimum achievable strain reaches a flat envelope (\(\mathcal F' \sim 0\)) away from the divergent points, the length coefficient vanishes (\(c_l \to 0\)) and the velocity coefficient approaches \(c_v \to 2\). Because \(|c_v|\) remains of order unity and the mass coefficient is unchanged, the order of magnitude of the required parametric precision does not change significantly across the frequency bands. Consequently, the low-frequency regime, where the coefficients are largest, can be considered the most stringent case. Requiring each parameter, taken individually, to contribute less than \(1\%\) to the phase error imposes
	\begin{equation}
		\frac{\Delta m}{m}\lesssim10^{-2},\qquad
		\frac{\Delta l}{l}\lesssim2.5\times10^{-3},\qquad
		\frac{\Delta v}{v}\lesssim5\times10^{-3}.
		\label{requirements}
	\end{equation}
	
	Table~\ref{tab:params} compares these requirements with the precision achieved or proposed in matter-wave interferometry. Quadrupole mass filters in Talbot-Lau interferometers have reached \(\Delta m/m=1/3000\)~\cite{haslinger2013universal}, comfortably below the \(10^{-2}\) requirement.  Velocity selectors reach \(\Delta v/v\simeq3\%\)~\cite{haslinger2013universal} and \(5\) to \(7\%\)~\cite{pedalino2026probing}. A continuous beam monochromator based on atom-surface diffraction has been proposed to reach fractional velocity spreads of order \(\Delta v/v \sim 10^{-3}\) for the velocity classes used in the massive-interferometry configuration considered here \cite{fiedler2024continuous}, which would meet the \(5\times10^{-3}\) requirement. The loop length of a Talbot-Lau interferometer is stabilized well below a micrometer, far exceeding the \(2.5\,\mathrm{mm}\) tolerance ~\cite{pedalino2026probing}. These capabilities indicate that the parametric requirements are compatible with existing technology. Instrumental imperfections (beam-splitter imbalance, detector efficiency, and trigger jitter) are not captured by Eq.~(\ref{deltaparam}) and require a separate analysis, deferred to future work.

	\begin{table}[h]
		\centering
		\footnotesize
		\renewcommand{\arraystretch}{1.4} 
		\caption{Parametric coefficients and precisions. The column ``Required'' gives the fractional precision yielding a \(1\%\) phase error [Eq.~(\ref{requirements})]. The last column quotes recent matter-wave values~\cite{haslinger2013universal,pedalino2026probing,fiedler2024continuous}.}
		\label{tab:params}
		\begin{tabular}{lcccc}
			\hline\hline
			Parameter & Coefficient & Low-\(\xi\) & Required & Achieved or proposed \\
			\hline
			Mass \(m\)     & \(1\)                          & \(1\)  & \(10^{-2}\)          & \(\ll 1\%\) \\
		Length \(l\)   & \(\xi\mathcal F'/\mathcal F\)  & \(4\)  & \(2.5\times10^{-3}\) & \(\Delta l\ll1\,\mu\mathrm{m}\) \\
		Speed \(v\) & \(2-\xi\mathcal F'/\mathcal F\) & \(-2\) & \(5\times10^{-3}\)   & \(3\) to \(7\%\), \(0.1\%\) proposed \\
			\hline\hline
		\end{tabular}
	\end{table}

	The massive-particle interferometry considered here differs fundamentally in its phase-accumulation mechanism from conventional laser interferometers (such as LIGO) and from a broad range of atom-interferometer proposals (such as light-pulse and material-grating schemes) developed for gravitational-wave detection. In detectors like LIGO, the GW signal is inferred from the differential optical phase associated with changes in the light propagation between suspended test-mass mirrors. In atom-interferometer proposals, the GW primarily alters an external reference field or structure (e.g., a laser baseline or a physical grating). The atomic ensembles then act as secondary phase sensors to read out the phase shift accumulated by these external components. Consequently, such schemes are limited by laser frequency noise, optical displacement noise, and platform vibrations, which must be carefully suppressed \cite{dimopoulos2009gravitational, geiger2017future, hogan2016atom}.
	
	By contrast, the present proposal models the GW-induced observable as the relative phase accumulated by a massive particle along the total path (e.g., a four-segment loop) in the interferometer. The GW contribution is calculated from the time integral of the effective tidal potential along the particle trajectory. Since the scheme does not require a laser link propagating along a long GW-detection baseline, it avoids the specific noises arising from the baseline laser-propagation phase contribution used in light-pulse gradiometer schemes. Its frequency response can also be tuned by changing the particle speed and the loop dimensions. These features identify a potentially complementary measurement architecture. Moreover, because the effect of mirror motion due to thermal noise is negligible for the total path, these interferometers may prove to be more robust against some important error sources that affect other types of interferometers.
	
	Nevertheless, important challenges remain, including maintaining coherence of the massive-particle superposition, achieving reliable splitting and recombination, and controlling velocity spread, environmental gravity gradients, seismic noise, and readout noise. These effects require further study, and future work may identify ways to mitigate them. Until they are fully quantified, the present results should be regarded as describing an idealized GW detector, one with the potential to access a broad range of GW frequencies and therefore worth pursuing further.
	
	\section*{Summary and conclusion}
	\label{S4}
	
	In summary, gravitational-wave observations offer a powerful window into fundamental physics, enabling precision tests of general relativity as well as studies of black holes and the early universe. The first direct detections marked a major milestone in modern physics while simultaneously underscoring the need for complementary detector concepts that can extend the accessible frequency range beyond that of existing observatories such as LIGO.
	
	Motivated by this need, we have investigated massive-particle interferometry as a potential platform for gravitational-wave detection. We showed that the interaction of a gravitational wave with a quantum particle propagating along spatially separated paths can generate a measurable phase shift, thereby opening a route to interferometric detection with massive quantum systems. In particular, we analyzed a massive-particle interferometric detector and examined how its sensitivity depends on particle mass, interferometer size, particle speed, and gravitational-wave frequency. Our results indicate that, within this idealized massive-particle interferometry, such a detector could in principle provide useful sensitivity over a broad frequency range, extending below the LIGO band, in a tabletop-scale apparatus.
	
	Although significant experimental challenges remain, rapid advances in matter-wave interferometry, macroscopic quantum superpositions, and the coherent control of massive particles render this approach a promising direction for future research. Carefully engineered massive-particle interferometers may therefore provide a complementary pathway toward next-generation gravitational-wave observatories.
	
	\section*{Acknowledgements}
	The author sincerely thanks Sohrab Rahvar, Shant Baghram, and Yusef Maleki for their enlightening discussions, careful and thoughtful reading of the manuscript, impactful comments, and many helpful discussions throughout this work. The author is also grateful to the anonymous reviewer for valuable and constructive feedback during the review process.
	
	\section*{Author contributions statement}
	A.M. conceived the study, developed the theoretical framework, performed the analytical derivations and numerical simulations, analyzed the results, and wrote the manuscript. All authors reviewed the manuscript.
	
	\section*{Funding}
	The author received no specific funding for this work.
	
	\section*{Competing interests}
	The author declares no competing interests.
	
	\section*{Data availability}
	The LIGO observatory data used to generate Figure 4 are available from the Gravitational Wave Open Science Center at \texttt{https://dcc.ligo.org/LIGO-P2000020/public}, as cited in Ref.~\cite{LIGO_P2000020}. All other data supporting the findings of this study are available within the article.

	\bibliographystyle{apsrev}
	\noindent \textbf{References}
	\bibliography{library}
	
\end{document}